\documentclass[a4paper,fleqn,usenatbib]{mnras}

\usepackage[T1]{fontenc}
\usepackage{ae,aecompl}

\usepackage{graphicx}	
\usepackage{amsmath}	
\usepackage{amssymb}	

\usepackage{color}

\title[Partially chaotic orbits]{Partially chaotic orbits in a perturbed cubic
force model}

\author[J. C. Muzzio]
{J. C. Muzzio$^{1,2}$,\thanks{E-mail: jcmuzzio@fcaglp.unlp.edu.ar}\\
$^{1}$Facultad de Ciencias Astron\'omicas y Geof\'isicas,
Universidad Nacional de La Plata, La Plata, Argentina\\
$^{2}$Instituto de Astrof\'isica de La Plata (CCT CONICET La Plata--UNLP), La
Plata, Argentina}

\date{Accepted XXX. Received YYY; in original form ZZZ}

\pubyear{2017}

\begin{document}
\label{firstpage}
\pagerange{\pageref{firstpage}--\pageref{lastpage}}
\maketitle

\begin{abstract}

Three types of orbits are theoretically possible in autonomous Hamiltonian
systems with three degrees of freedom: fully chaotic (they only obey the
energy integral), partially chaotic (they obey an additional isolating
integral besides energy) and regular (they obey two isolating integrals
besides energy). The existence of partially chaotic orbits has been denied
by several authors, however, arguing either that there is a sudden transition
from regularity to full chaoticity, or that a long enough follow up of a
supposedly partially chaotic orbit would reveal a fully chaotic nature. This
situation needs clarification, because partially chaotic orbits might play a
significant role in the process of chaotic diffusion. Here
we use numerically computed Lyapunov exponents to explore the phase
space of a perturbed three dimensional cubic force toy model, and a
generalization of the Poincar\'e maps to show that partially chaotic orbits
are actually present in that model. They turn out to be double orbits joined
by a bifurcation zone, which is the most likely source of their chaos, and
they are encapsulated in  regions of phase space bounded by regular orbits
similar to each one of the components of the double orbit.

\end{abstract}




\section{Introduction}

In autonomous Hamiltonian systems of three degrees of fredom regular
orbits obey two isolating integrals besides energy and, in principle,
chaotic orbits may either conserve only the energy or obey also an
additional integral. In fact, \citet{Cont1978} found in one of those
systems three regions where the motions obeyed,
respectively, two integrals besides energy, one integral besides energy
or only the energy integral. The existence of at least three disjoint
invariant regions with a different 'degree of stochasticity' on the same
energy surface of two Hamiltonian systems (one of them that of
\citet{Cont1978}) was also reported by \citet{PV1984} who, from the very
title of their paper, noted that this fact would imply a possible failure
of Arnold diffusion 

Several studies of the dynamics of triaxial stellar systems noticed the
presence of two types of chaotic orbits \citep[see, e.g.,][]{GS1981, MV1996},
but no importance was assigned then to recognizing one from the other.
Nevertheless \citet{M2003} reported that, in a strongly triaxial model
elliptical galaxy, the orbits with only one positive Lyapunov exponent had
a spatial distribution different from that of orbits with two positive
exponents, dubbing them partially and fully chaotic orbits, respectively.
He identified the former with the orbits that have one additional integral
besides energy and the latter with those that obey the energy integral only.
His results were confirmed by \citet{MM2004} and \citet{MCW2005} and the
distinction between partially and fully chaotic orbits was included in all
the subsequent work on triaxial stellar systems by him and his co-workers
\citep[see][and references therein]{CM2016}. It should be
noted that their partially chaotic and fully chaotic orbits are just
the same as, respectively, the weakly chaotic and strongly chaotic orbits
of \citet{PV1984}. The change of wording was justified by \citet{MCW2005}
because the terms weak and strong chaos had been used in connection with the
maximum value of the Lyapunov exponent \citep[][]{C2002}.

The phenomenon of 'stable chaos' in the Solar System investigated by
\citet{MN1992} and \citet{Mil1997} might be another astronomical example of
partially chaotic motion. Although those authors only computed the largest
Lyapunov exponent for the cases they considered, the fact that the size of
their chaotic regions in phase space is small suggests that possibility. 

Nevertheless, \citet{F1970a} investigated a case of the restricted three-body
problem and found that, when varying one parameter, the two isolating integrals
other than energy
seemed to vanish at the same time, rather than one after the other. Later on,
using a four-dimensional mapping, \citet{F1971} concluded that a system with
three degrees of freedom has in general either none or two isolating integrals
besides energy, and that the dissapearance of one of those integrals entails
the dissapearance of the other.

\citet{LL1992} discussed the Lyapunov exponents computed by \citet{Ben1980}
for the Hamiltonian system investigated by \citet{Cont1978} and raised an
important point. They argued that, due to Arnold diffusion, an orbit initially
in the region where \citet{Cont1978} found partially chaotic orbits would end
up in their region of fully chaotic orbits after a sufficiently long time.
Thus, since Lyapunov exponents are computed over finite time intervals, an
orbit with only one positive exponent might turn out to have two positive
exponents if integrated over a long enough interval, and partially chaotic
orbits found with this and similar methods might actually be fully chaotic
orbits in disguise. They concluded that such methods should be supplemented
by other techniques in order to clarify the true nature of the chaotic motion.
Clearly, any proof obtained with numerical methods that
cover a certain time interval can be regarded as valid only over that interval,
and its extension to infinite time can only be obtained by analytical methods.
From a practical point of view, however, a numerical proof that covers an
interval much longer than the lifetime of the system investigated may
be all that is needed (e.g., many Hubble times for galactic astronomy).

Since few authors have distinguished partially from fully chaotic
orbits, the phenomenon of stickiness \citep[see, e.g.,][]{SR1982, M1985,CH2010}
is usually described as the behaviour of a chaotic orbit (either partially or
fully chaotic) that remains for long intervals close to regions of regularity
mimicking a regular orbit and then moves into chaotic regions showing its true nature.
To that phenomenon we should add now the one indicated by \citet{LL1992}, i.e.,
a fully chaotic orbit can remain for a long time behaving as a partially chaotic
one and, later on, display its fully chaotic character. Thus, a sticky orbit can be
not only a chaotic orbit that behaves as regular for a long time, but also a
fully chaotic orbit that behaves as partially chaotic for a long time.

As noted by \citet{PV1984} and \citet{LL1992} the existence, or not, of
partially chaotic orbits and Arnold diffusion are closely related, so that
we will analyze that relationship in more detail. Ultimately, the basis of
Arnold diffusion is the fact that a region of N dimensions (N-D hereafter)
can be partitioned in disjoint subregions only by objects of (N-1)-D. For
example, we can partition a surface (2-D) using curves (1-D), but not
points (0-D). And we can partition a 3-D volume using surfaces (2-D), but
not curves (1-D) or points (0-D). Now, in an autonomous Hamiltonian system
with two degrees of freedom, chaotic motion takes place in a 3-D space (the
4-D phase space minus one dimension due to the energy integral) and the
tori of the regular orbits are 2-D (another dimension is subtracted by the
additional integral), so that chaotic orbits are constrained to regions
bounded by those tori. But for autonomous systems with three degrees of
freedom the situation is very different. Their phase space is 6-D, so that
fully chaotic orbits are 5-D, partially chaotic orbits are 4-D and regular
orbits are 3-D. Clearly, the 3-D orbits cannot prevent the 5-D fully
chaotic orbits from moving through all the space allowed to them by the
energy integral (except for the 3-D space occupied by the regular orbits
themselves) and that is Arnold diffusion in a nutshell (a caveat is that
the regular 3-D tori should still remain, when they broke we have
resonance superposition rather Arnold diffusion). Such scenario is clearly
altered if partially chaotic orbits actually exist because, in that case,
the 3-D regular orbits could limit the motion of the 4-D partially chaotic
orbits which, in turn, would place barriers to the motion of the 5-D fully
chaotic orbits.

The present paper is just a first step to try to establish whether that
situation is possible. Here we show that partially chaotic orbits are
present in a toy model and that they are bounded by regular orbits.
We computed the six Lyapunov exponents of orbits and,
besides, we used cuts and slices in the 5-D space to generalize the usual
Poincar\'e maps of the 3-D space. We also made extensive use of 3-D plots using
colour as the fourth dimension, a technique developed by \citet{PZ1994} and
extensively used later on by them and their coworkers
\citep[see, e.g.,][]{KP2011, Kats2011, Kats2013}, including studies of
the relative location of chaotic orbits with respect to tori
in 3-D Hamiltonian systems. Colour and slices were also used by \citet{Rich2014}
to investigate 4-D symplectic maps.
 
Our methods cannot be easily extended to check whether fully chaotic orbits
are, in turn, bounded by partially chaotic orbits and that will be the
subject of a future investigation.

The following section describes our model and the numerical techniques we
used to study its orbits. The results of a search for possible partially
chaotic orbits using Lyapunov exponents are presented in Section 3, where
we also use the concept of the second integral to isolate one of them and,
in Section 4, we show that it is actually a partially
chaotic orbit bounded by regular orbits. Finally, our conclusions are
presented in Section 5.

\section{Model and numerical methods}

\subsection{The model}
\label{model} 

We chose the perturbed cubic force model whose Hamiltonian is:
\begin{equation}
    H =\frac{(u^2+v^2+w^2)}{2}+\frac{x^4+y^4+z^4}{4}+\epsilon x^2(y+z),
	\label{hamilt}
\end{equation}
where $x, y$ and $z$ are the coordinates, $u, v$ and $w$ the
corresponding impulses and $\epsilon$ regulates the size of the perturbation.
For the present investigation we adopted $H=1$ and $\epsilon=0.005$.

This model was thoroughly investigated by \citet{Cinc2003} and we refer
the reader to that paper for full details. Here we will only mention a few
properties of the model that are important for the present investigation.

The unperturbed system (i.e., $\epsilon = 0$) conserves the energies in each
coordinate:
\begin{equation}
    h_{1}=\frac{u^2}{2}+\frac{x^4}{4},
	\label{energy1}
\end{equation}
\begin{equation}
    h_{2}=\frac{v^2}{2}+\frac{y^4}{4},
	\label{energy2}
\end{equation}
\begin{equation}
    h_{3}=\frac{w^2}{2}+\frac{z^4}{4},
	\label{energy3}
\end{equation}
and the unperturbed frequency in each coordinate is proportional to
the fourth root of the corresponding energy ($\omega_{i} \propto h_{i}^{1/4}$).
The unperturbed resonances:
\begin{equation}
   l \times \omega_{1} + m \times \omega_{2} + n \times \omega_{3} = 0,
	\label{reson}
\end{equation}
with $l, m$ and $n$ integers and not all three equal to zero, form the
Arnold web. Hereafter we will refer to such resonances as $(l,m,n)$.

It is convenient to adopt the following coordinates:
\begin{equation}
    e_{1}=\frac{h_{1}-2h_{2}+h_{3}}{\sqrt{6}},
	\label{e1}
\end{equation}
\begin{equation}
    e_{2}=\frac{h_{1}-h_{3}}{\sqrt{2}},
	\label{e2}
\end{equation}
\begin{equation}
    e_{3}=\frac{h_{1}+h_{2}+h_{3}}{\sqrt{3}},
	\label{e3}
\end{equation}
that have the advantage that $e_{1}$ and $e_{2}$ lie in the unperturbed
energy plane and $e_{3}$ is perpendicular to it. \citet{Cinc2003} present
several figures showing the resonant structure in the $(e_{1}, e_{2})$ plane,
and the distribution of MEGNO levels (their chaoticity indicator) on that
plane. Similar figures had also been presented by \citet{Froe2000} for a
different model. There one can see that resonances are the 'threads', and
double resonances \footnote {Actually, if two resonant conditions are
fullfilled, any linear combination of them is fullfilled too, so that we have
an infinite number of resonances. For simplicity, however, we will refer to
that condition as a double resonance.} the 'knots', of the Arnold web. Double
resonances, in particular, present central regions occupied by regular orbits
surrounded by chaotic areas, so that they look as a promising
hunting ground for partially chaotic orbits bounded by regular orbits.

In the present investigation we will deal mainly with the double resonance
$(2,-1,0)$ and $(0,1,-1)$ and we will see that the projections of orbits on
the $(e_{1}, e_{2})$ plane are more or less symmetric with respect to a line
parallel to the projection of the latter resonance on that plane. Therefore,
it will be useful to also use the coordinates:
\begin{equation}
    e_{p}=e_{1} cos(\alpha) + e_{2} sin(\alpha), 
	\label{ep}
\end{equation}
\begin{equation}
    e_{n}=-e_{1} sin(\alpha) + e_{2} cos(\alpha),
	\label{en}
\end{equation}
with $\alpha = 60$ degrees.

\subsection{Numerical methods}
\label{methods} 

We explored the phase space of our model using orbits with initial coordinates
$x=y=z=0$ and different initial velocity components (all positive) such that
$u^2+v^2+w^2=2$, in order to have $H=1$.
In the region of the double resonance mentioned above
the interval between two crossings of the $x=0$ plane is of the order of $12.6$
time units ($t.u.$ hereafter), which can be taken as a characteristic time of
our investigation. To follow each orbit and at the same time compute the
six Lyapunov exponents (LEs hereafter) we used the LIAMAG routine, kindly
provided by D. Pfenniger \citep[see][]{UP1988}. However, we replaced the original
Runge-Kutta-Fehlberg subroutine by the high order Taylor subroutine of
\citet{JZ2005} (subroutine and documentation can be obtained at
http://www.maia.ub.es/$\sim$angel/taylor/) which allowed us to perform about 50
times longer integrations, with the same precision as the original subroutine.
The LEs $\lambda_{1} > \lambda_{2} >...> \lambda_{5} > \lambda_{6}$  have the
property that $\lambda_{i}=-\lambda_{7-i}$, due to the conservation of the volume
in phase space, and that $\lambda_{3}=\lambda_{4}=0$, due to the conservation of
the energy integral. Besides, each additional isolating integral makes zero another
$\lambda_{i}=-\lambda_{7-i}$ pair so that, considering only the three largest
LEs, we have that all three are zero for regular orbits, only two are zero
for partially cahotic orbits, and only the third one is zero for fully chaotic
orbits.

Nevertheless, the properties of the numerically computed LEs differ somewhat
from those described above, because theoretical LEs are defined
for an infinite time interval while the orbit integrations that allow their
numerical computation are necessarily finite. Therefore, numerical LEs can tend
towards zero as the integration time increases, but they remain always larger
than a limiting value that can be estimated to be of the order of $ln T/T$,
where $T$ is the length of the integration interval. This is a coarse estimate
only, and the limiting value should be determined in every case
\citep[see, e.g.,][for details]{ZM2012}. In the present investigation we used
integration times of $5 \times 10^{6}$, $5 \times 10^{7}$, $5 \times 10^{8}$
and $5 \times 10^{9}$ $t.u.$ and energy conservation was better than about
$0.5\times10^{-12}$, $1.5\times10^{-12}$ $5.\times10^{-12}$ and $20.\times10^{-12}$,
respectively. Notice that $5 \times 10^{9}$ $t.u.$ corresponds
to $4 \times 10^{8}$ characteristic times in the present investigation, while 1 Hubble
time corresponds to about 200 characteristic times for the elliptical galaxies we
investigated previously \citep[see, e.g.,][]{ZM2012}, so that our longest integrations
cover intervals equivalent to $2 \times 10^{6}$ Hubble times in a galactic context.

\begin{figure}
	\includegraphics[width=\columnwidth]{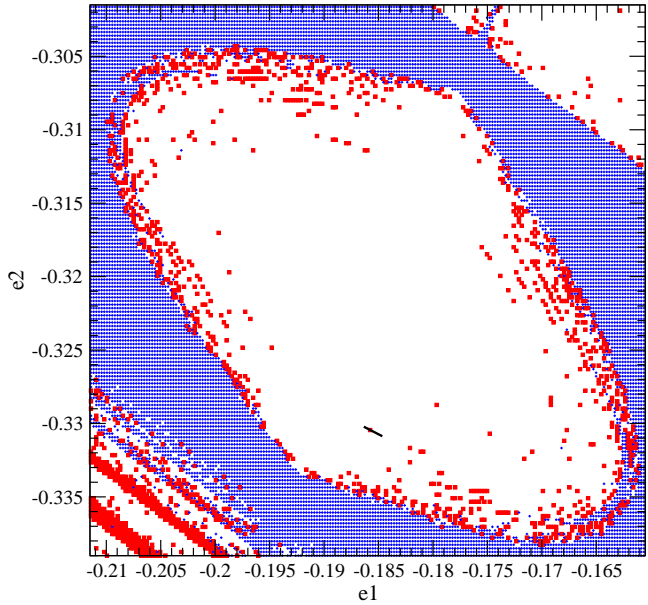}
    \caption{Initial conditions on the $(e_{1}, e_{2})$ plane of
orbits classified as regular, partially and fully chaotic from
the values of their LEs near the double resonance $(2,-1,0)$ and $(0,1,-1)$.
The blank areas correspond to regular orbits, partially chaotic orbits are
shown as filled squares (red in the electronic version) and fully chaotic
orbits as plus signs (blue in the electronic version). The
short black line shows the region covered by Figure~\ref{fig02}.}
    \label{fig01}
\end{figure}

A powerful method for the investigation of autonomous Hamiltonian systems with
two degrees of freedom is offered by the Poincar\'e maps. In those systems chaotic
orbits are 3-D and regular orbits 2-D so that if we make a cut, e.g. requesting
that $x=0$ (with $u > 0$ to avoid the sign indetermination) and plot the
resulting points on the $(y, v)$ plane, chaotic orbits
appear as surfaces and regular orbits as lines. It is tempting to
try to extend this method to systems with three degrees of fredom where fully
chaotic orbits are 5-D, partially chaotic orbits 4-D and regular orbits 3-D: with
two cuts we could obtain plots where partially chaotic orbits would appear as
surfaces and regular orbits as lines. But there are several complications on such
a naive extension of the method. To begin with, while one can make a very precise
first cut using the method of \citet{H1982}, the second cut can only be done
approximately (e.g., taking $\vert y \vert \leq 0.00010$), and that is why
the method has been dubbed 'slice cutting' by
\citet{F1970b}. But as the width of the slice is reduced to get more
accuracy in the plots, the number of points is drastically reduced so that very
long integrations of the orbit are necessary to get a reasonable number of
accurate points. Besides, the two cuts do not generally give a plane, as
happens with the original Poincar\'e method, but a warped surface, because
it is embbeded in a 3-D space rather than in the 2-D space of the original
method. Finally, in order to compare orbits, one has to take orbits that not
only have the same energy (as in the usual 2-D Poincar\'e maps) but also the
same value of the second integral, and we have no mathematical expresion
for that integral. Therefore, the graphical representation and interpretation of
these plots, that we will call hereafter 3-D Poincar\'e maps, are much less
straightforward than those of the original method.

\begin{figure*}
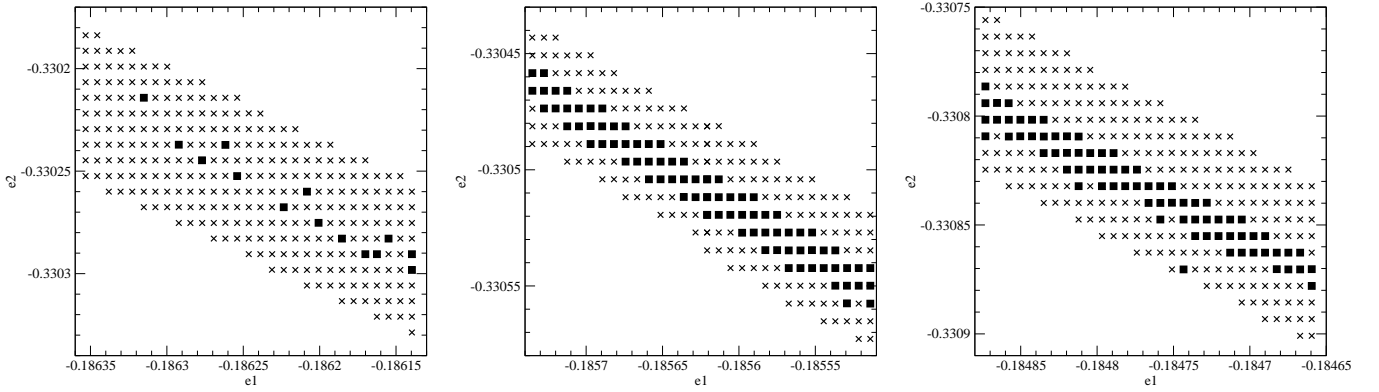

\resizebox{\hsize}{!}{\includegraphics{laneleft.eps}\hspace{1cm}
                      \includegraphics{lanecenter.eps}\hspace{1cm}
                      \includegraphics{laneright.eps}}
    \caption{High resolution plots of small sections of Figure 1. Regular orbits
are shown as crosses and partially chaotic orbits as filled squares.}
    \label{fig02}
\end{figure*}

Despite those difficulties, we thought that the 3-D Poincar\'e maps might
offer a useful tool to prove whether partially chaotic orbits can be bounded
by regular orbits. We resorted again to the high order Taylor routine of
\citet{JZ2005} to make a program that allowed us to follow orbits for a
long time, to do a precise first cut using the method of \citet{H1982}
and a second, less precise, cut simply selecting a small range of values
for another variable. Most of the results presented here correspond to a
first cut taking $x=0$ (with $u > 0$), a second cut with
$\vert y \vert \leq 0.00010$ (plus $v > 0$) and a total integration time of
about $1\times10^{9}t.u.$. Energy conservation for our 3-D Poincar\'e maps
was better than $1.3\times10^{-11}$. The $1\times10^{9}t.u.$
interval corresponds to $8\times10^{7}$ characteristic times of our system
so that, repeating the comparison we did above for the integration time of
the LEs, we may conclude that it is equivalent to about $4\times10^{5}$
Hubble times in a galactic context.

\begin{figure}
	\includegraphics[width=\columnwidth]{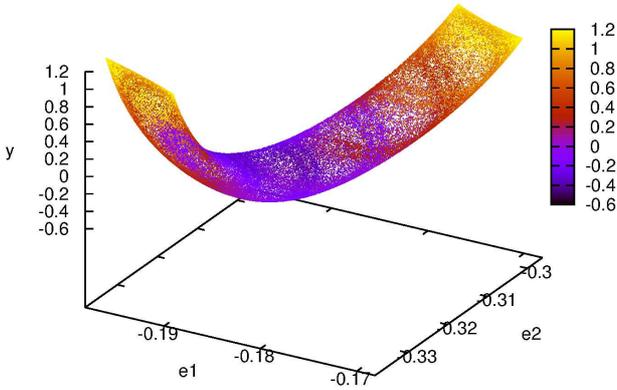}
    \caption{A cut with the plane $x=0$ ($u>0$) of a partially chaotic
orbit from the chain shown in Fig. 2. It is shown in the 3-D space
$(e_{1}, e_{2}, y)$ with the fourth dimension $z$ given by the colour
scale in the electronic version.}
    \label{fig03}
\end{figure}

It was very important at several stages of our investigation to get a clear
view of orbital structures and, for that purpose,
we found very useful the technique of \citep[][]{PZ1994} who used 3-D plots
plus colour to represent the fourth dimension. We have adopted their method
using gnuplot (Copyright (C) 1986 - 1993, 1998, 2004, 2007 Thomas Williams,
Colin Kelley) to make the plots.

As mentioned above, in 3-D the cuts yield warped surfaces and not planes, so
that the graphic representation of the Poincar\'e maps is quite a challenge.
But for our purposes the problem is simplified, because what we need to
show is that the surfaces that represent the partially chaotic orbits are
bounded by the lines that represent the regular orbits. Nevertheless, the
separation between those orbits is very small compared with the size of the
orbits themselves, so that a simple plot on, say, the $(e_{1}, e_{2})$ plane
will lose all the details. After careful examination of the orbits on
different projections and on 3-D plots using colour as the fourth dimension
\citep[][]{PZ1994}, we adopted the following method. We chose the $e_{1}$,
$e_{2}$ and $z$ variables and, for an orbit selected as reference, we normalized
each one of those variables subtracting the corresponding value for the center of the orbit and
dividing the result by the dispersion of the variable in question. Then we transformed that
normalized coordinate system into a spherical one
and, using the azimuth angle $\phi$ as argument, we obtained
the best fitting Fourier series for the polar angle $\theta$ and the
radius $r$. For nearby orbits, we obtained the differences
between the values of their parameters (normalized using the same center and
dispersions of the orbit taken as reference) and those given by the corresponding
Fourier series, so that we have $\phi$ as a 'long' variable and the differences
between their $\theta$ and $r$ values and the corresponding ones of the orbit taken as
reference can be plotted with considerable detail.

\begin{figure}
	\includegraphics[width=\columnwidth]{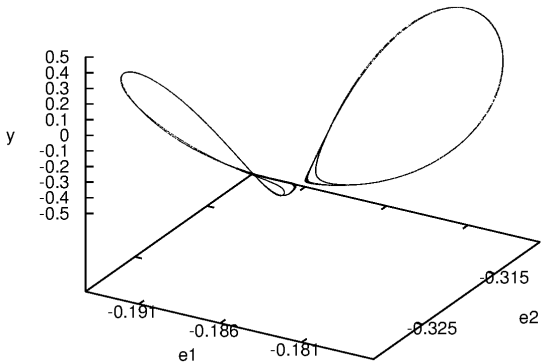}
    \caption{A slice $\vert y \vert \leq 0.00010$ (v>0) of the cut of
Fig. 3 in the 3-D space $(e_{1}, e_{2}, z)$.}
    \label{fig04}
\end{figure}

We experimented with different numbers of terms and
found that the mean square error decreased as we increased that number up to
about 121 terms (that is, up to terms $sin(60\phi)$ and $cos(60\phi)$) and
reached a plateau where increasing the number of terms did not significantly
decreased the mean square error any further, so that we adopted that number
of terms for our computations. For slices with $\vert y \vert \leq 0.00010$
the resulting mean square errors of the $e_{1}$, $e_{2}$ and $z$ variables
turned to be of the order of $0.3 \times 10^{-5}$, $0.5 \times 10^{-5}$ and
$0.4 \times 10^{-4}$, respectively, which imply errors relative to
the range of the corresponding variable of the order of $0.03$, $0.04$ and
$0.005$ per cent, respectively. For slices with $\vert y \vert \leq 0.00020$
and $\vert y \vert \leq 0.00005$ the errors were twice larger and one-half
smaller, respectively, i.e., proportional to the width of the slice as
could be expected. To estimate the errors of integration we obtained the
Fourier series using only the first 20 per cent points and computed the
mean square errors of the last 20 per cent points with respect to those series.
The dispersions turned out to be essentially the same, so that
the errors of integration should be much smaller than the
dispersion caused by the slice widths.

\section{Partially chaotic orbits}

The first step of our investigation was to search for possible partially
chaotic orbits in regions of the phase space populated mainly by regular
orbits and, as indicated above, we chose the region of the double
resonance $(2,-1,0)$ and $(0,1,-1)$. We performed our search using LEs,
so that the already mentioned warning of \citet{LL1992} applies: we can
never be sure that LEs obtained with longer integrations would not reveal
that the partially chaotic orbits found in that way are actually fully
chaotic orbits on disguise. Thus, it should be recalled
that the orbits that we will refer to as partially chaotic in the present
section can be regarded as such only over time intervals of the order of those
covered by our numerical integrations.

\subsection{The search}
\label{energyplane}

The $(2,-1,0)$ and $(0,1,-1)$ resonances of the unperturbed Hamiltonian
(equation~{\ref{hamilt}, with $\epsilon = 0$) cross on the $(e_{1}, e_{2})$
plane at $(-5/(11\sqrt{6}, -5/(11\sqrt{2}) \simeq (-0.18557, -0.32141)$.
Therefore, we prepared a sample of initial conditions taking $H=1$, $x=y=z=0$ and
a grid of $e_{1}$ and $e_{2}$ values, with $2^{-12} \simeq 2.44 \times 10^{-4}$
spacing, centered at that point. The advantage of taking these
initial conditions is that the energy of our full Hamiltonian is the same as that of
the unperturbed Hamiltonian there. Besides, as we will take cuts with $x=0$ later on,
the energies of the unperturbed and perturbed Hamiltonians at those cuts will
again be the same, i.e., the value of the coordinate $e_{3}$ will be conserved. With
those initial conditions, and using the full Hamiltonian with
$\epsilon = 0.005$, we computed the  orbits and obtained the LEs with an
integration time of $5 \times 10^{6} t.u.$, which we used to
classify the orbits as regular, partially or fully chaotic.
Our results are presented in Figure~\ref{fig01}, where we note a central region
dominated by regular orbits, surrounded by another one dominated by fully chaotic
ones, with most of the partially chaotic orbits lying on the border between
those regions. That border is by no means clear cut and displays considerable
structure but, what interests us here is that many partially chaotic orbits
appear also well inside the regular domain and they even seem to form chains
there. In all likelihood they correspond to resonances.

Three sectors of one of those chains are shown on Figure~\ref{fig02}, which
was obtained in the same way as Figure~\ref{fig01}, but with a much finer grid
spacing of $2^{-17} \simeq 7.63 \times 10^{-6}$. The same
region is shown by a short black line on Figure~\ref{fig01}. The chain extends
continuously in between the sectors shown and beyond the third one,
but it seems on the verge of dissapearance on the first sector and probably
it does not extend much further in that direction. Several of the partially
chaotic orbits found here were followed with integrations of $5 \times 10^{7} t.u.$,
$5 \times 10^{8} t.u.$ and $5 \times 10^{9} t.u$ and their LEs were computed,
and in all cases they continued to appear as partially chaotic.

Figure~\ref{fig03} shows, in the 3-D space $(e_{1}, e_{2}, y)$ and using colour
to represent the fourth dimension $z$, the cut $x=0~(u>0)$ of the orbit
whose initial condition lies on $e_{1}=-0.18490365$ and $e_{2}=-0.33078644$
and that we will dub pch hereafter;
it was obtained with an integration over $1. \times 10^{10} t.u.$. This is one
of the partially chaotic orbits that lie on the chain shown in Figure~\ref{fig02}
and the form of the cut reminds that of a croissant. Some mixing of the
colours might be present, a characteristic of chaotic orbits in this sort of
plot as indicated by \citet{PZ1994} but, if it exists, it is far from clear. In
fact, except perhaps for the colour distribution, similar plots for other orbits
from the same region, either regular or partially chaotic, look very much the same.
Thus, while Figure~\ref{fig03} is useful to show us the general aspect of these
orbits, we need more precise ways to proceed with our investigation. As indicated
above, the croissant is approximately symmetric with respect to a plane normal to
to the $(e_{1}, e_{2})$ plane and parallel to the line drawn on the latter plane
by the $(0,1,-1)$ resonance.

Figure~\ref{fig04} shows the result of taking a slice $\vert y \vert \leq 0.00010~(v>0)$
from the $x$ cut of Figure~\ref{fig03} in the 3-D space $(e_{1}, e_{2}, z)$. As
anticipated, the points lie on a warped surface (actually, it has a very small
width because there is a finite range of $y$ values) and not on a plane. There are
two separate lobes which correspond to the two slices taken from the croissant by
the slice $\vert y \vert \leq 0.00010$ (compare with Figure~\ref{fig03}). For the
time being, we will concentrate on the lobe that corresponds to the larger values
of $e_{1}$ and $e_{2}$.

Figure~\ref{fig05} shows the projections of that lobe on the
$(e_{1}, e_{2})$ plane, for the same partially chaotic orbit of the previous two
Figures, and for two regular orbits with initial conditions $e_{1}=-0.184892701$,
$e_{2}=-0.33077844$ (r1 hereafter) and $e_{1}=-0.18490851$ and $e_{2}=-0.33079272$
(r2 hereafter), respectively, which lie each one on each side of the partially
chaotic lane of Figure~\ref{fig02}. We notice that the partially chaotic orbit is
double and, as we will see later, it is not just a double loop but the two parts
cross themselves, i.e., it bifurcates and that is the cause of its chaoticity. One
can already note that the outer tip of the orbit near $e_{1}=-0.1852$ and $e_{2}=-0.3218$
is thicker than the neighbouring inner tip, and the reason is that the former is a
surface and not a line, as we will see later on. Interestingly, each one of the
two regular orbits is similar to each one of the two parts of the regular orbit.
In other words, the lane of partially chaotic orbits shown in Figure~\ref{fig02} 
is a transition zone from regular orbits similar to r1 to regular orbits similar
to r2. Thus, it seems reasonable to assume that the 4-D partially chaotic orbits
of that lane are bounded by the 3-D tori of the regular orbits that border the
lane and, in terms of Figure~\ref{fig03}, we could expect the 3-D croissant of a
partially chaotic orbit to be bounded by 2-D croissants of regular orbits. Let us
explore how to check that possibility.

\begin{figure}
\centering
\includegraphics[width=7.6cm]{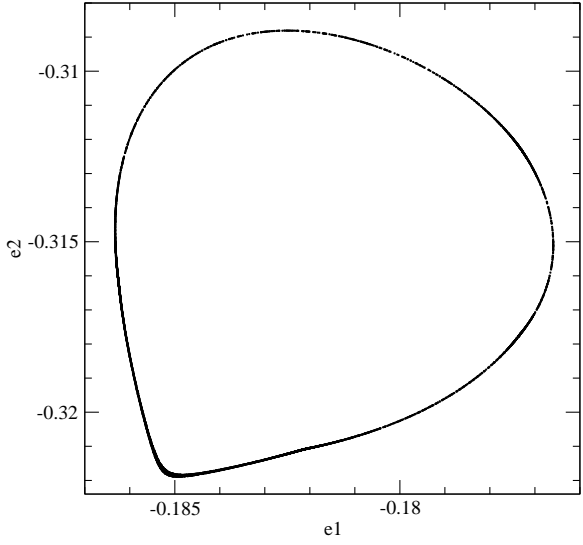}
\vskip 5.mm
\includegraphics[width=7.6cm]{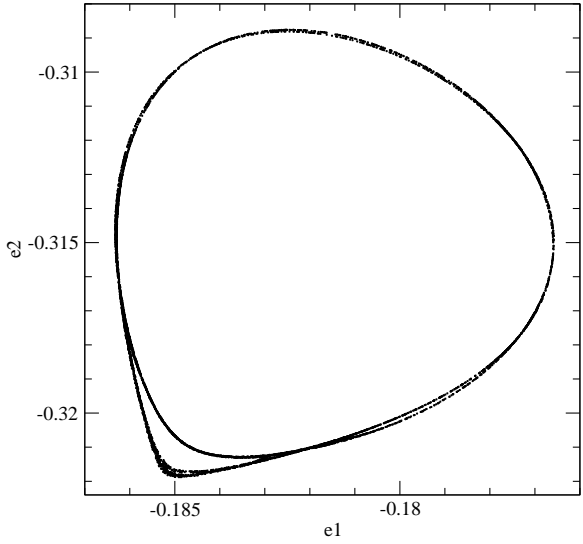}
\vskip 5.mm
\includegraphics[width=7.6cm]{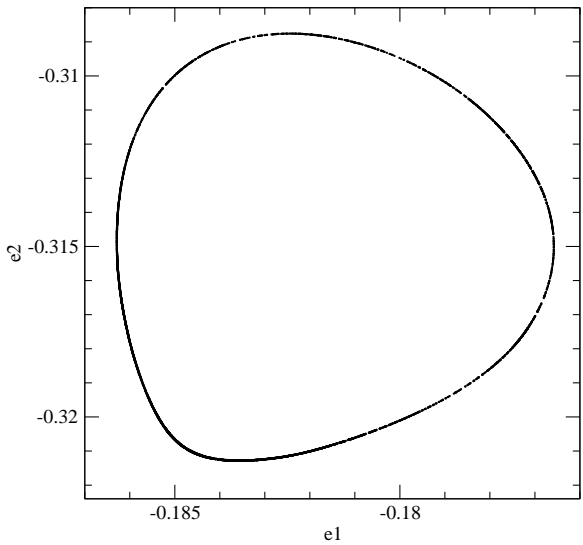}
    \caption{Projections on the $(e_{1}, e_{2})$ plane of the cuts $x=0~(u>0)$
and $\vert y \vert \leq 0.00010~(v>0)$ of the orbits r1 (above), pch (center)
and r2 (below).}
    \label{fig05}
\end{figure}

\subsection{The second integral of motion}
\label{integral}

As we mentioned above, for our 3-D Poincar\'e maps we have to select orbits with
the same value of the energy and the same value of the second integral, but the
problem is that we do not know the mathematical expresion of that second integral.
Since the initial conditions we used to map the unperturbed energy plane had
all the same energy and $x = y = z = 0$ Figure~\ref{fig02} is, except for the
fact that it includes orbits with different values of the second integral, a
kind of Poincar\'e map resulting from the three cuts $x = 0$, $y = 0$ and
$z = 0$, so that every partially chaotic orbit should appear there are as a
1-D curve. Therefore, the lane of partially chaotic orbits in Figure~\ref{fig02}
can be seen as the surface that results from the superposition of the curves
corresponding to partially chaotic orbits with different values of the second
integral. Our problem is, thus, to identify the individual curves that make up
that surface. A simple idea would be to follow a single partially chaotic orbit
and to take a third cut, or slice, $z \simeq 0$ in addition to $x = 0$ and
$y \simeq 0$, but it fails because it is impossible get a reasonable number
of points with sufficient accuracy: if one uses small widths for the $y$ and
$z$ slices, one gets almost no points, and if one increases those widths to
get enough points, they appear distributed on a surface rather than on a line. 
Another simple idea, to take the points that result from the $x = 0$ and
$y \simeq 0$ cuts and lie near $z = 0$, to fit them to a surface and to
determine the intersection of that surface with the plane $z = 0$ also
fails for similar reasons, plus the fact that the surface in
question is warped and difficult to adjust. Finally, we reasoned that the
points in Figure~\ref{fig02} are all very accurate, because the initial
conditions of the corresponding orbits had been selected by ourselves,
so that what we had to do was to determine which points of the partially
chaotic lane corresponded to the same orbit, i.e., we had to find which
initial conditions gave the same distribution of points resulting from
the $x = 0$ and $y \simeq 0$ cuts. 

The first step to that approach was to find a suitable sector of the
partially chaotic orbits to do the comparison and the tip near
$e_{1}=-0.1852$ and $e_{2}=-0.3218$ (see Figure~\ref{fig05}) was an
obvious choice. In 3-D plots in the $(e_{p}, e_{n}, z)$ space (see
equations~\ref{ep} and \ref{en}) it looks like a half cilinder with
its axis lying on the  $(e_{n} = -0.00055, z)$ plane. The $z$ height
of its minimum turned out to be highly dependent on the orbit in
question, so that we decided to take a slice of $0.00010$ width around
$e_{n} = -0.00055$, plus similar slices around $e_{n} = -0.00045$ and
$-0.00035$ as additional aid. 

\begin{figure}
\centering
\includegraphics[width=7.6cm]{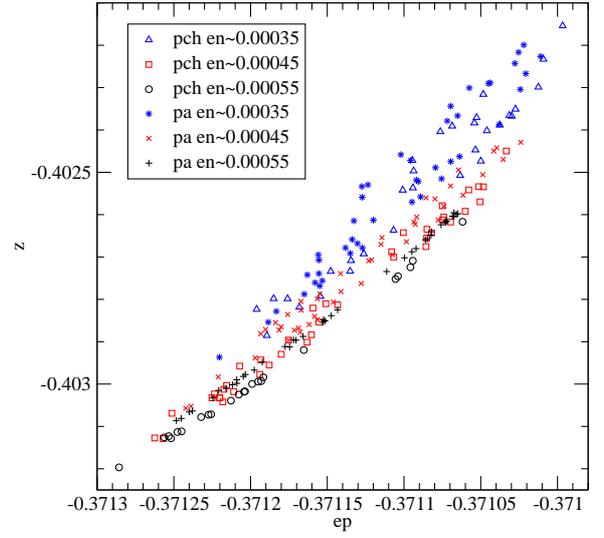}
\vskip 5.mm
\includegraphics[width=7.6cm]{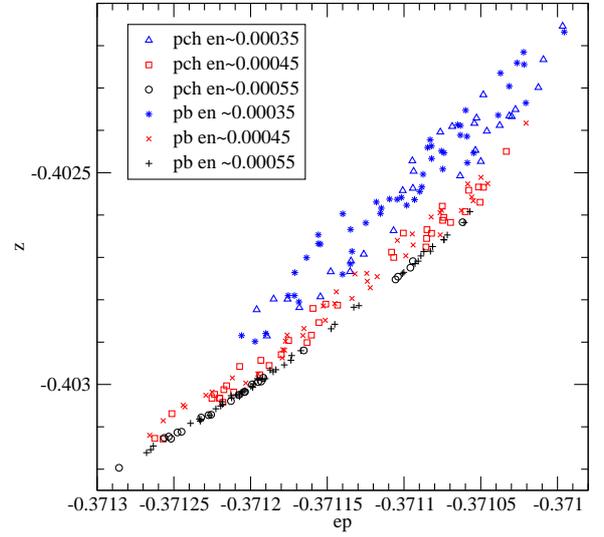}
\vskip 5.mm
\includegraphics[width=7.6cm]{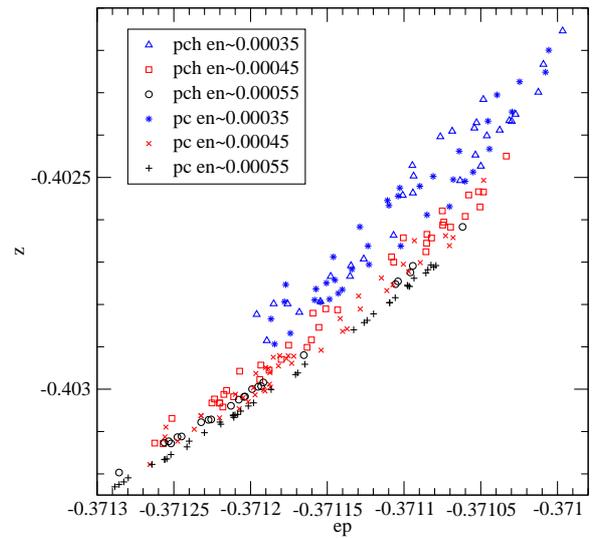}
    \caption{Projections on the $(e_{p}, z)$ plane of slices around
$e_{n} = -0.00055, -0.00045$ (red in the electronic version) and $-0.00035$
(blue in the electronic version) of the cuts $x=0~(u>0)$ and
$\vert y \vert \leq 0.00010~(v>0)$  of the orbit pch, together with
those of the orbits pa (above), pb (center) and pc (below).}
    \label{fig06}
\end{figure}

Figure~\ref{fig06} shows the results for the partially chaotic orbit
pch from Figures~\ref{fig03}, 
\ref{fig04} and~\ref{fig05} together with three other partially chaotic orbits
from the lane shown on Figures~\ref{fig02}: pa (above) with initial conditions
at $e_{1} = -0.18489818$ and $e_{2} = -0.33078078$, pb (center) with initial
conditions at $e_{1} = -0.18489736$ and $e_{2} = -0.33078126$ and pc (below)
with initial conditions at $e_{1} = -0.18489653$ and $e_{2} = -0.33078174$.
Despite the very small differences among the initial conditions, it is clear
that orbits pa and pc are not the same as orbit pch, while orbit pb coincides
perfectly well with it. Figure~\ref{fig07} shows the projections of orbits
pch and pb on the $(e_{p}, e_{n})$ plane confirming the excellent fit.

\begin{figure}
\resizebox{\hsize}{!}{\includegraphics{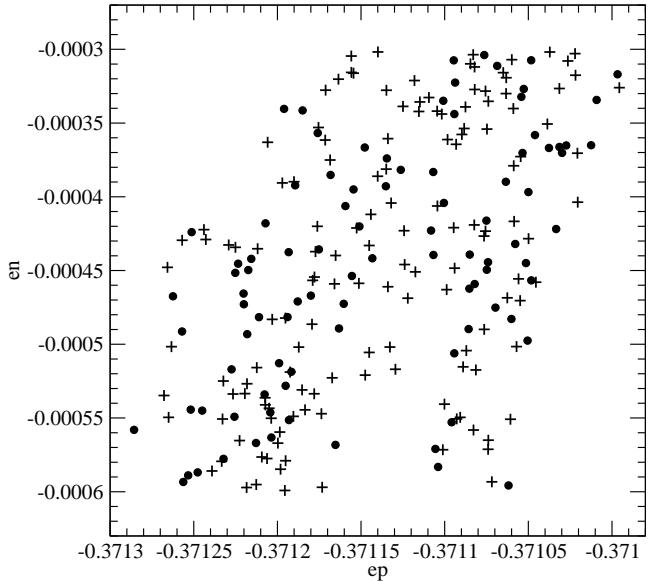}}
    \caption{Projections on the $(e_{p}, e_{n})$ plane of the cuts $x=0~(u>0)$
and $\vert y \vert \leq 0.00010~(v>0)$ of the orbits pch (filled circles) and
pb (plus signs).}
    \label{fig07}
\end{figure}

Then we prepared a plot of the initial conditions on the $(e_{1}, e_{2})$ plane
that produce regular and partially chaotic orbits, similar to those of
Figure~\ref{fig02} but with a finer grid spacing of
$2^{-19} \simeq 1.91 \times 10^{-6}$ and with the grid tilted 60
degrees to follow better the direction of the partially chaotic lane. It is
shown in Figure~\ref{fig08} with blank spaces for the regular orbits and filled
small circles for the partially chaotic ones. Taking the results of orbit pch
as reference, we prepared plots similar to those of Figure~\ref{fig06} for
partially chaotic orbits
with the initial conditions of nearby points on the $(e_{1}, e_{2})$ plane
and used them to select the initial conditions that provided good fits. In
several instances the points from the grid did not give a good enough fit and
we used additional points interpolated between those that gave results above and
below those of orbit pch. It was hard work, indeed, because for each
$(e_{1}, e_{2})$ point we had to compute the orbit and the $x = 0, y \simeq 0$ cuts,
do the plots and examine them by eye, but the results were excellent and are shown
as big open circles on Figure~\ref{fig08}. Those open circles correspond all to one
single partially chaotic orbit and, therefore, they trace the line along which
the second integral is constant.
 
\begin{figure}
\resizebox{\hsize}{!}{\includegraphics{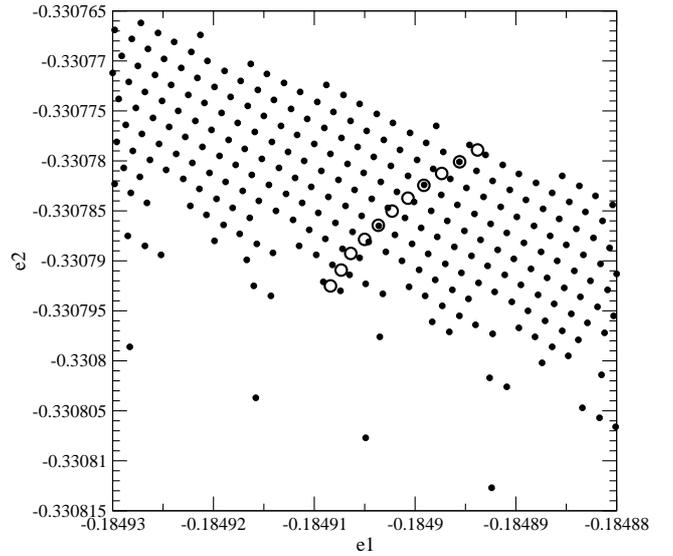}}
    \caption{Initial conditions on the $(e_{1}, e_{2})$ plane of orbits
subsequently clasified as regular (blank space) or partially chaotic (small
filled circles). The large open circles correspond all to the same partially
chaotic orbit, see the text for explanation.}
    \label{fig08}
\end{figure}

\section{Bounding regular orbits}
\subsection{Finding the boundaries}
\label{boundaries}
 
It seems natural to try to extend the method we used to find the trace of
the second integral of a partially chaotic orbit to search for the regular
orbits that have the same value of that integral and, therefore,
bound that orbit. For that purpose, we selected from our grid
on the $(e_{1}, e_{2})$ plane points that corresponded to regular orbits
and were located near the extrapolation of the line we had found for the
second integral. Taking again the partially chaotic orbit pch as reference,
we used plots like the one shown in Figure~\ref{fig09} to find which
regular orbits offered the best extrapolation to the results of pch. In
that way we selected the orbit r1 shown in that Figure (see also
Figure~\ref{fig05}). Of course, since the tori of the regular orbits that
have the same values of the energy and of the second integral fit one
inside the other like Russian dolls many can be found and
orbit r1 is just one reasonable choice among other possible ones.
 
\begin{figure}
\resizebox{\hsize}{!}{\includegraphics{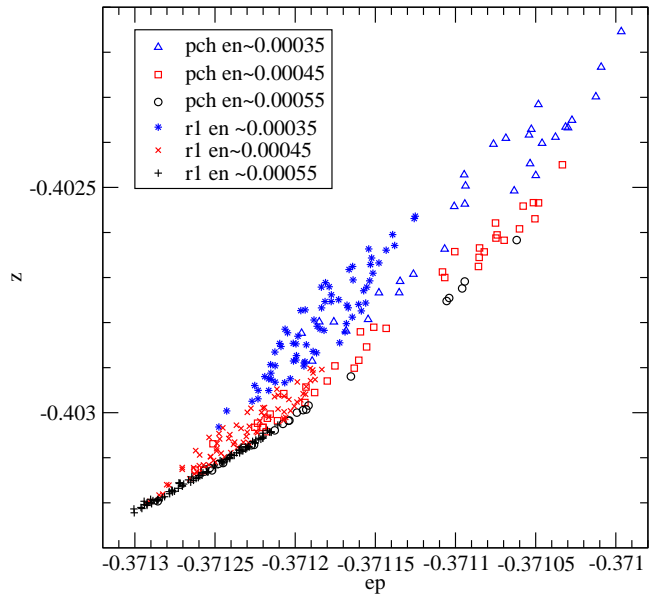}}
    \caption{Same as Figure 6, but for partially chaotic orbit pch and
regular orbit r1.}
    \label{fig09}
\end{figure}

Figure~\ref{fig10} shows the projection of the same orbits pch and r1
on the $(e_{p}, e_{n})$ plane and one can notice again the good fit.
The upper part of the Figure was obtained with a slice
$\vert y \vert \leq 0.00010~(v>0)$, as used for all the previous
Figures, and the lower part with a slice $\vert y \vert \leq 0.00005~(v>0)$.
Clearly, the reduced slice width reduces almost in half the width of the
band occupied by the regular orbit, as could be expected since for $y = 0$
we should get a line. But the global width of the region occupied by the
partially chaotic orbit is only slightly reduced, as should happen because
for $y = 0$ we should get a surface. Besides, we note that for the regular
orbit the points with $y < 0$ are clearly separated from those with $y > 0$,
the former lying towards the left and the latter towards the right, further
proof that the width of the band of r1 points is essentially due to the
width of the $y$ slice. Moreover, we note that for the partially chaotic
orbit the points with positive and negative values of $y$ appear mixed, as
they should because such an orbit lacks the second isolating integral besides
energy that has the regular orbit. Nevertheless, the effect of the width of
the $y$ slice can also be noticed for orbit pch because its points that lie
on the extreme left have $y < 0$ and those that lie on the extreme right
have $y > 0$. Last but not least, it is clear that the apparent overlap of
the r1 and pch orbits on Figures~\ref{fig09} and \ref{fig10} is only apparent
and just caused by the thickness of the $y$ slice.

\begin{figure}
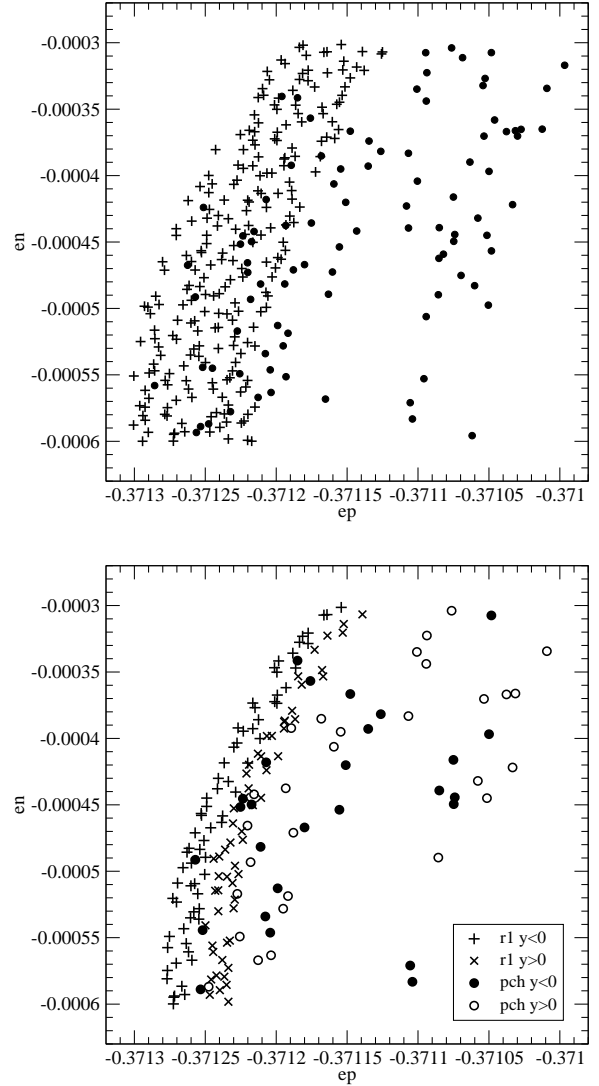

\centering
\includegraphics[width=7.6cm]{epenpchr1.eps}
\vskip 5.mm
\includegraphics[width=7.6cm]{epenpchr1pr.eps}
    \caption{Projections on the $(e_{p}, e_{n})$ plane of the cuts $x=0~(u>0)$
and $\vert y \vert \leq 0.00010~(v>0)$ (above) and $\vert y \vert \leq 0.00005~(v>0)$
(below)  of the orbits r1 (crosses) and pch (circles).}
    \label{fig10}
\end{figure}

Unfortunately one cannot apply the same procedure to find the bounding regular
orbits below the partially chaotic lane because regular orbits there lack
the tip near $e_{1}=-0.1852$ and $e_{2}=-0.3218$ (see Figure~\ref{fig05}),
so that we had to look for another region of the partially chaotic orbit
where the $x$ and $y$ cuts yielded a suitable surface. The bifurcation regions
seemed an obvious choice and Figure~\ref{fig11} shows one of those regions
of orbit pc together with the adjacent regions of regular orbits r1 and r2.
We had found r1 as described above, and we found r2 searching for another
good boundary in the way we describe below. The plot presents the cut
$x = 0$ (u>0) and $\vert y \vert \leq 0.00010~(v>0)$ in the
$(e_{1}, e_{2}, z)$ space and using colour for the fourth dimension $y$.
We note that, as indicated by \citet{PZ1994}, the colours appear mixed on
the surface that corresponds to the partially chaotic orbit because, since
one integral of motion is lacking, $y$ is not correlated to the other
variables. For the regular orbits, instead, there is a clear progresion
from negative values of $y$ at left and below to positive values at right
and above, i.e., these orbits are, as they should, lines and their thickness
is only due to the width of the $y$ slice.
 
\begin{figure}
\resizebox{\hsize}{!}{\includegraphics{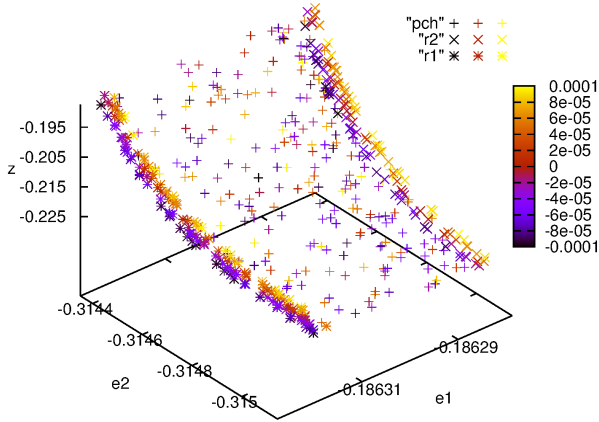}}
    \caption{One of the bifurcation zones of the partially chaotic orbit pch
together with the nearby sections of regular orbits r1 and r2.
We show the result of cuts $x = 0$ (u>0) and  $\vert y \vert \leq 0.00010~(v>0)$
in the $(e_{1}, e_{2}, z)$ space and using colour for the fourth dimension $y$
in the electronic version.}
    \label{fig11}
\end{figure}

The form of the partially chaotic surface resembles again that of a half
cilinder, so that we might proceed in a similar way as we
did to find the bounding orbit r1. But, in that case, the sector of the partially
chaotic surface was oriented in the $(e_{p}, e_{n}, z)$ space in a way that
facilitated our purposes. In the present case we need to find a system of
coordinates that allows us to see part of the cilinder edge on so as to be
able to select an adequate bounding orbit as the limit of that edge.
We found that three rotations were necessary:
1) One of an angle $\theta$ around the $z$ axis, to bring the $(e_{1}, e_{2}, z)$
system to a new $(d_{1}, d_{2}, z)$ system; 2) A second one of an angle $\psi$
around the $d_{1}$ axis to bring the latter to the $(g_{1}, g_{2}, d_{1})$
system; 3) A final one of an angle $\tau$ around the $g_{2}$ axis to obtain
the $(q_{1}, q_{2}, g_{2})$ system. We took the points of the orbit pch
shown in Figure~\ref{fig11} and, for different trial values of $\theta$,
we determined the $\psi$ and $\tau$ angles that aligned the region covered
by those points with the $d_{2}$ and $z$ axes and with the $g_{1}$ and
$d_{1}$ axes, respectively, and we computed the $z$ thickness of the $(d_{2},z)$
projection. Finally, we adopted the angles that minimized that thickness,
which turned out to be $\theta = 43.60, \psi = 89.21$ and $\tau = 89.25$
degrees. It is quite possible that other systems of coordinates
might be used for the same purpose, but the one described worked well and that
was all we needed here. To search for the regular orbit that bounded the partially chaotic
orbit pch we used plots on the $(q_{2}, g_{2})$ and the $(q_{2}, q_{1})$
planes; in the first case we plotted separately the points within the
slices $(-0.235 < q_{1} < -0.228)$, $(-0.228 < q_{1} < -0.221)$ and
$(-0.221 < q_{1} < -0.214)$, in a similar way as we had done to search
for the limiting regular orbit r1 on the other side of the partially
chaotic lane.
 
\begin{figure}
\resizebox{\hsize}{!}{\includegraphics{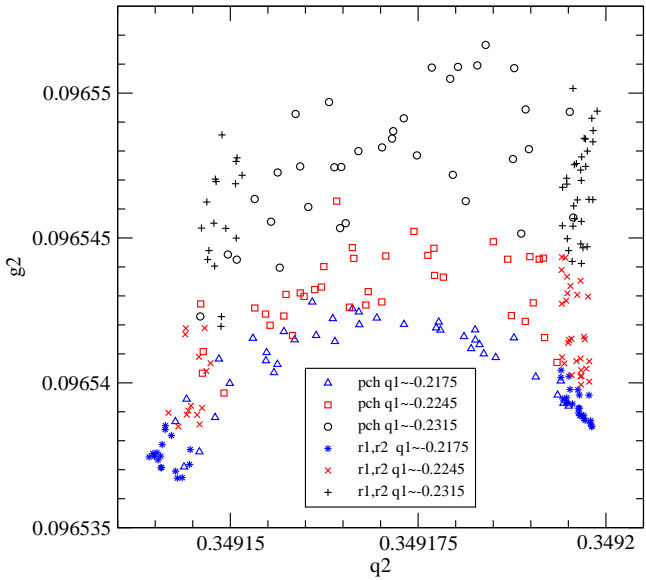}}
    \caption{Projection on the $(q_{2}, g_{2})$ plane of slices around
$q_{1} \simeq -0.2175$ (blue in the electronic version), $-0.2245$ (red
in the electronic version) and $-0.2315$ of the cuts $x = 0~(u > 0)$
and $\vert y \vert \leq 0.00010~(v>0)$ of the partially chaotic
orbit pch and of regular orbits r1 (right) and r2 (left).}
    \label{fig12}
\end{figure}

Figure~\ref{fig12} shows the diagram we used to select regular orbit r2
as the boundary to partially chaotic orbit pch and we have included also
orbit r1, which provides the other boundary. While r2 was found searching
for a good boundary to the surface shown in the Figure, r1 had been found
from fits to a completely different region of the partially chaotic orbit,
so that its excellent agreement with this new region gives a strong support
to our method. Figure~\ref{fig13} shows the projection on the $(q_{2}, q_{1})$
plane confirming our results.
 
\begin{figure}
\resizebox{\hsize}{!}{\includegraphics{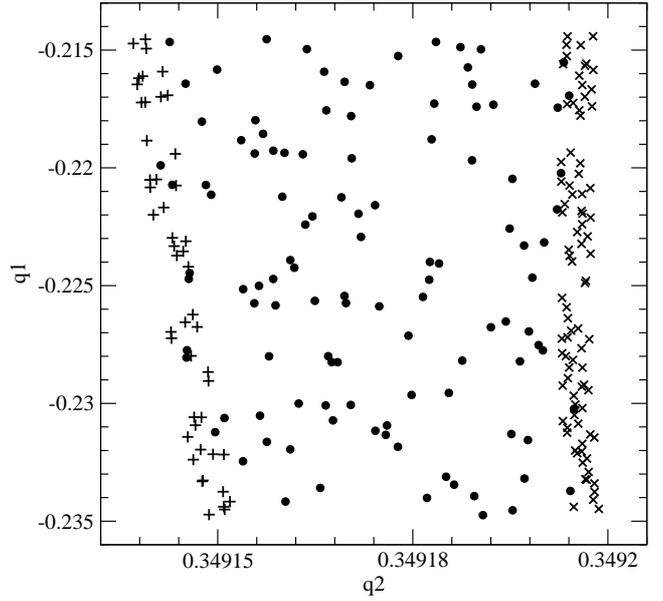}}
    \caption{Projection on the $(q_{2}, q_{1})$ plane of the cuts $x = 0~(u > 0)$
and $\vert y \vert \leq 0.00010~(v>0)$ of the partially chaotic orbit pch
(filled circles) and the regular orbits r1 (crosses) and r2 (plus signs).}
    \label{fig13}
\end{figure}

\subsection{3-D Poincar\'e maps}
\label{maps}

We adopted as our reference orbit the right upper lobe that results from
the cuts $x~(u > 0)$ and  $\vert y \vert \leq 0.00005~(v>0)$ of
orbit r2 and we computed the mean values ($<e_{1}>, <e_{2}>$
and $<z>$) and the dispersions ($\sigma_{1}, \sigma_{2}$ and $\sigma_{z}$), of
its $e_{1}$, $e_{2}$ and $z$ values. Adopting those mean values as the
center of the orbit, we computed the normalized values $(e_{1}-<e_{1}>)/\sigma_1,
(e_{1}-<e_{1}>)/\sigma_1$ and $(z-<z>)/\sigma_z$ and used these normalized
values to define a new spherical system of coordinates, with azimuth angle
$\phi$, polar angle $\theta$, and radius $r$. Then, taking $\phi$
as argument, we adjusted each normalized coordinate with a Fourier series
and we used them to iteratively improve the center of the orbit. Finally,
we obtained new Fourier series to represent $\theta$ and $r$
as functions of $\phi$. The differences between the
true values and those given by the series, i.e. the residuals, were
used to represent orbit r2 in our 3-D Poincar\'e maps, that is,
straight lines with some dispersion through $\theta = 0$ and $r = 0$,
respectively.

\begin{figure}
\centering
\includegraphics[width=7.6cm]{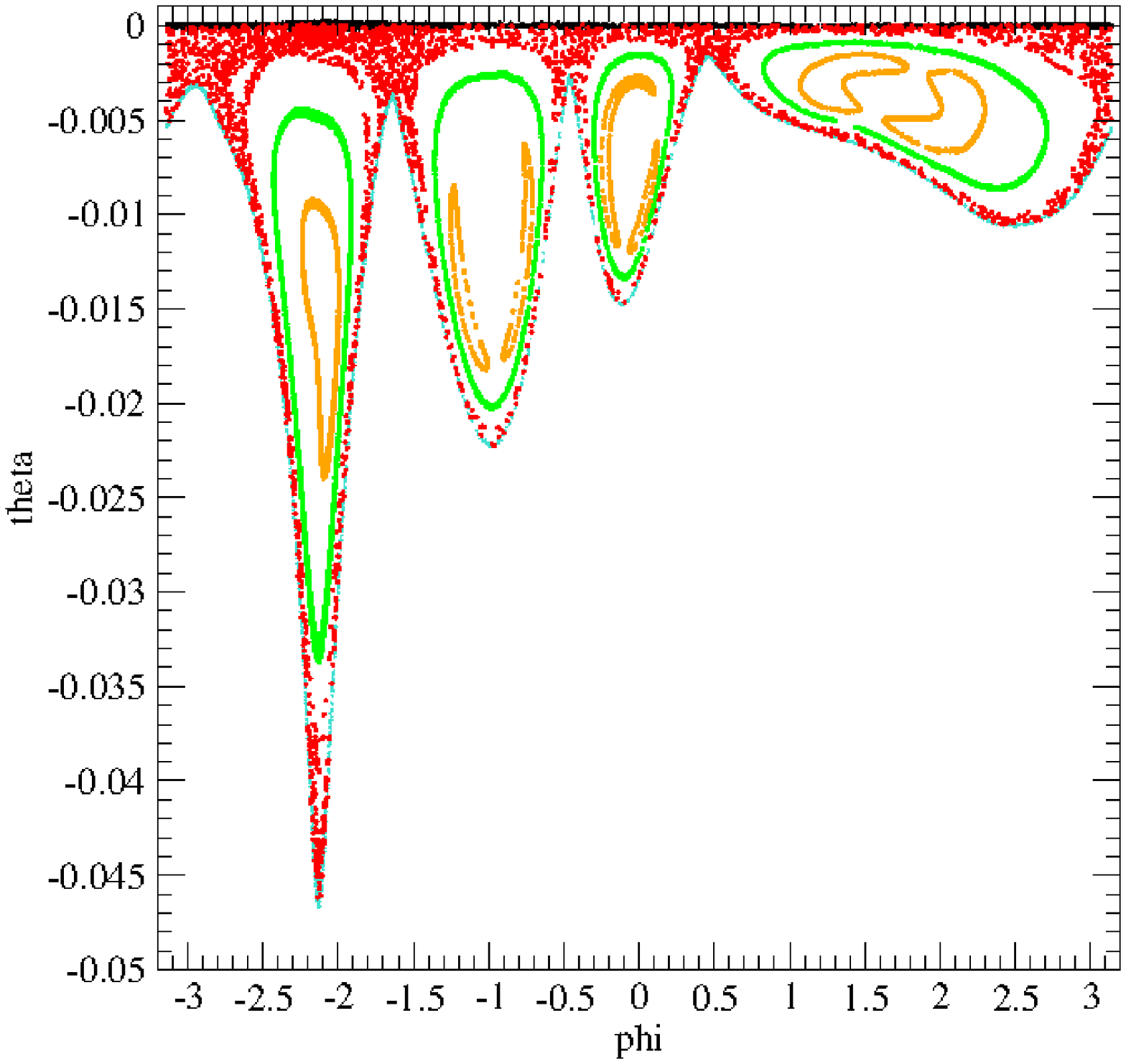}
\vskip 5.mm
\includegraphics[width=7.6cm]{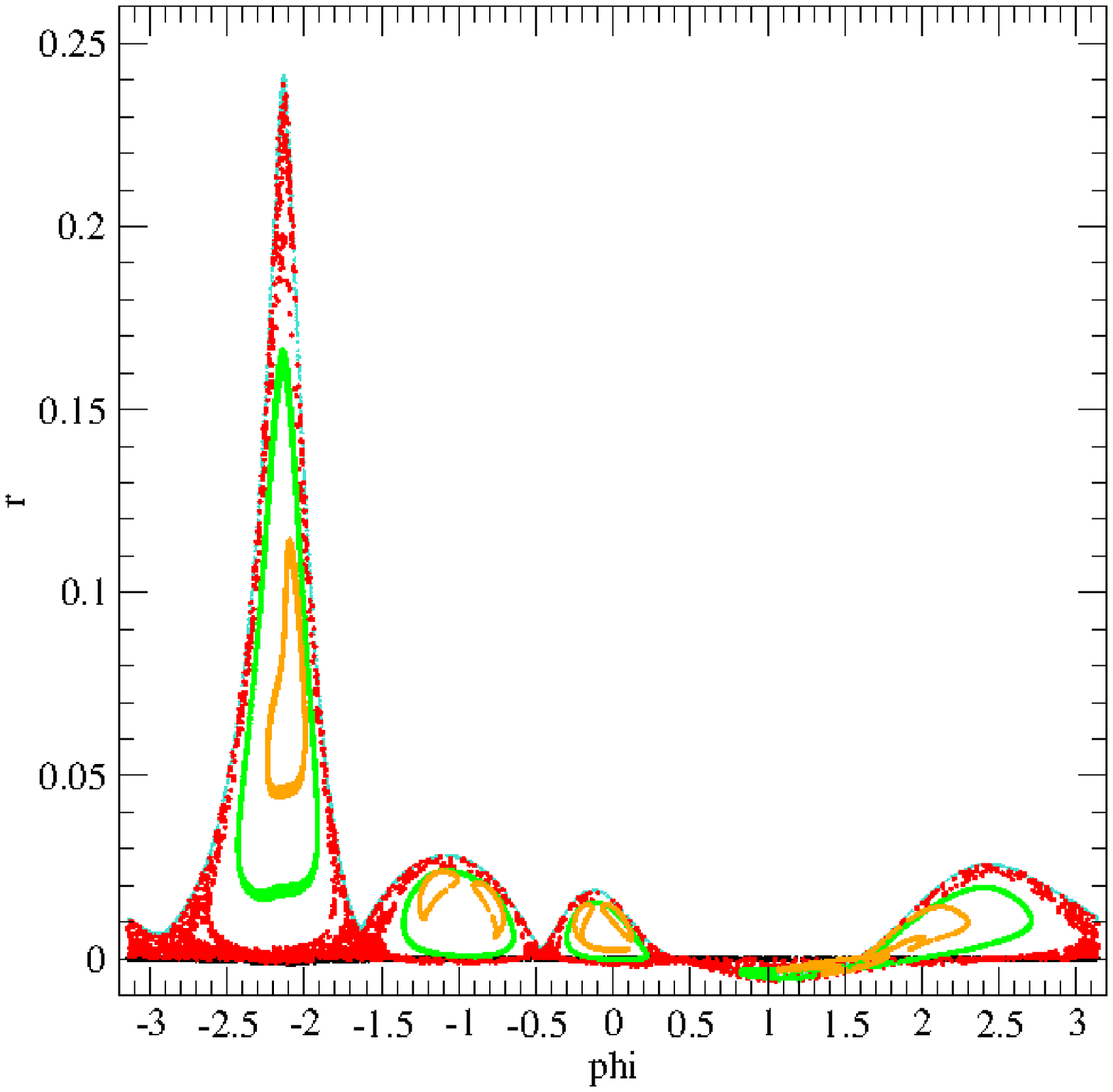}
    \caption{3-D Poincar\'e maps $(\phi, \theta)$ (above) and $(\phi, r)$
(below) of orbits r1 (turquoise in the electronic version), r2, pch (red in
the electronic version), r3 (green in the electronic version) and
r4 (orange in the electronic version). They correspond
to the $x = 0$ (u > 0) and $\vert y \vert \leq 0.00005~(v>0)$
cuts and to the right lobe. The ordinates give the differences between
the values of $\theta$ and $r$, respectively, of each orbit and those given by the
Fourier series fitted to the correponding values of orbit r2. See text for explanation.}
    \label{fig14}
\end{figure}

For other orbits we normalized their $e_{1}, e_{2}$ and
$z$ values using the same center and dispersions adopted for the r2 orbit
and obtained the corresponding $\phi, \theta$ and $r$ values.
Finally, using their $\phi$ values as argument of the Fourier series
obtained for r2, we obtained the differences between their $\theta$
and $r$ values and those given by the series. In other words,
our 3-D Poincar\'e maps are just the differences between each orbit
and r2, so that we can clearly represent those small differences as
we follow the orbit through all the different azimuth angles.

\begin{figure}
\centering
\includegraphics[width=7.6cm]{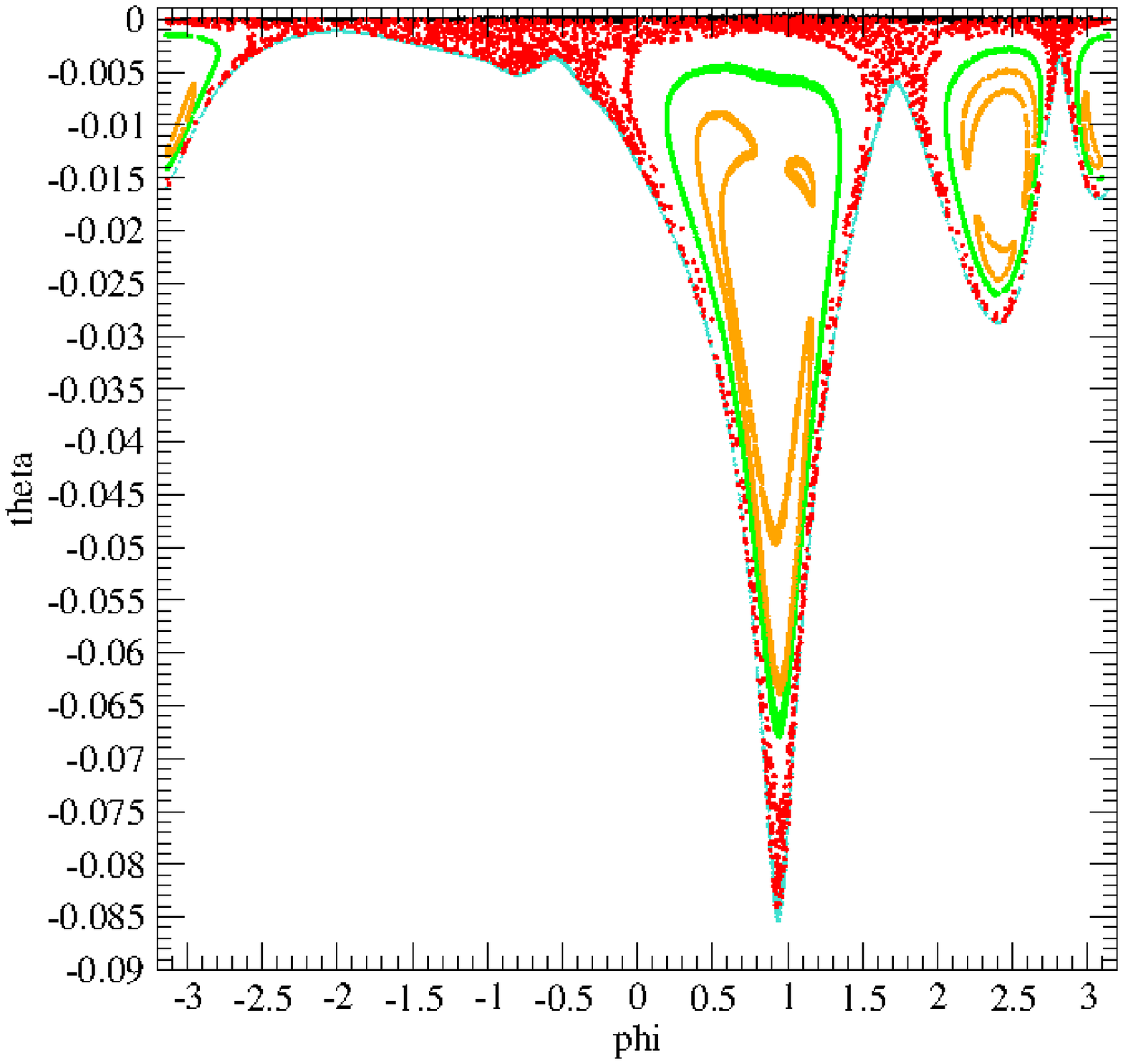}
\vskip 5.mm
\includegraphics[width=7.6cm]{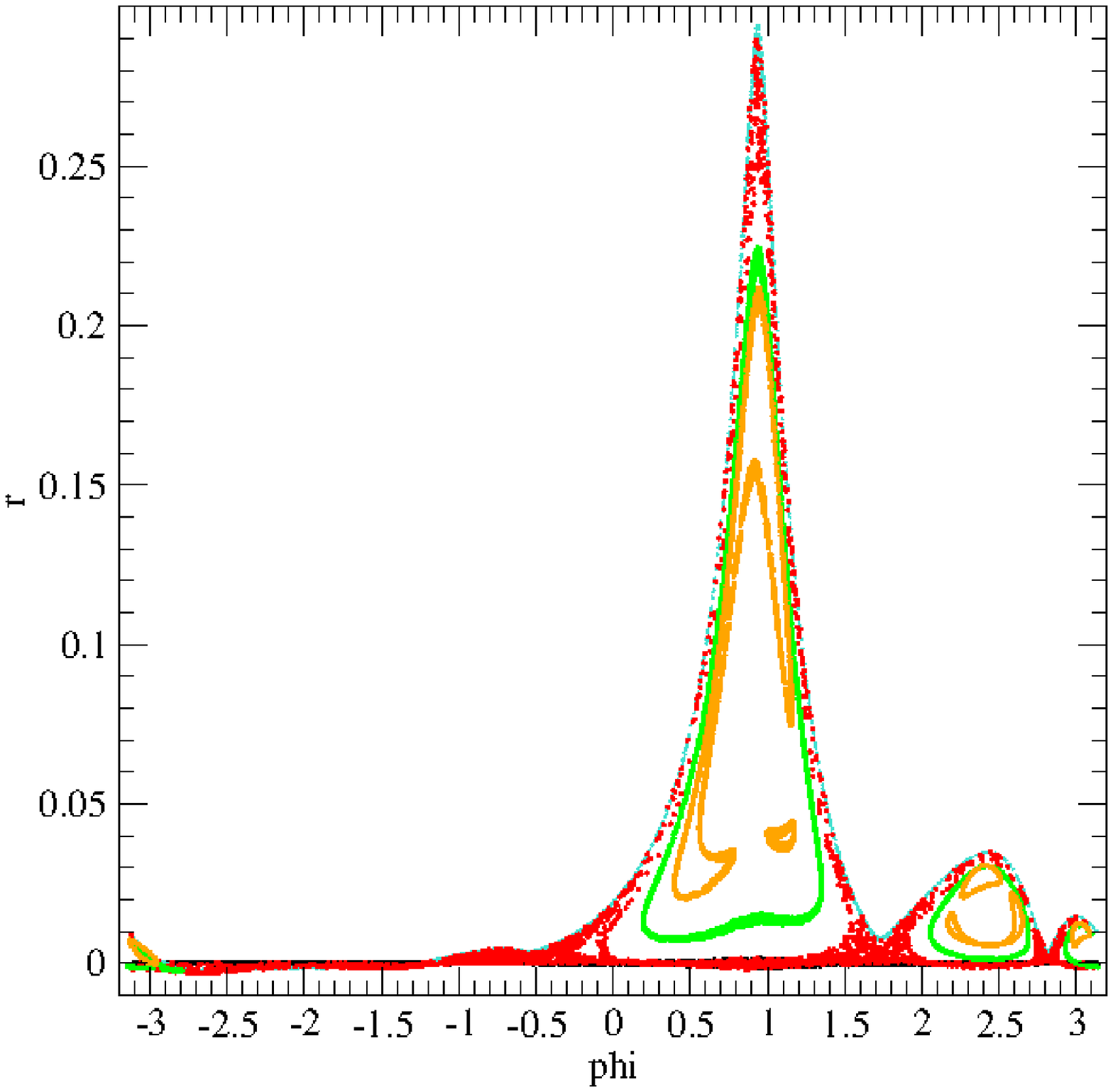}
    \caption{Same as Figure 14, but for the left lobe.}
    \label{fig15}
\end{figure}

Figure~\ref{fig14} presents the result. Besides the partially chaotic orbit
pch (red dots in the electronic version) and the bounding orbits r1 (turquoise dots
in the electronic version) and r2 (black dots), we included regular orbits r3
(green dots in the electronic version) and r4 (orange dots in the electronic
version) which lie in the main holes of the partially chaotic orbit. r3 and
r4 were found interpolating values within those holes, have initial condition $x = 0$, plus
$e_{1} = -0.18637000, e_{2} = -0.31992000, y = -0.39720000, z = 0$ for r3 and 
$e_{1} = -0.18528000, e_{2} = -0.32372000, y = 0, z = -0.39498000$ for r4.
As those orbits have velocities of opposite signs in alternate holes,
in this case the conditions $u > 0$, at the $x$ cut, and $v > 0$, at the
$y$ cut, were not applied. The fact that they are regular
orbits is supported not only because they appear as lines on the Poincar\'e map, but
also by computations of their LEs. We even found a regular orbit that lies in the
smaller holes of the partially chaotic orbit, with initial conditions
$x = 0, y = 0, z = -0.40282500, e_{1} = -0.18510000, e_{2} = -0.32168100$, but
we preferred not to include it in the Figure in order that those smaller
holes could be better seen. One should recall that we are dealing with
warped surfaces and not with planes, and that is the reason why some orbits
seem to cross on the $(\phi, r)$ plot, but there are no crossings on the
$(\phi, \theta)$ plot and, therefore, neither are there in space. Except for
that caveat, there are no significant differences between our plots,
particularly the $(\phi, \theta)$ plot, and a standard
Poincar\'e map for an autonomous Hamiltonian with two degrees of freedom: the
partially chaotic orbit pch occupies a surface and is bounded by the lines that
represent the regular orbits r1 and r2; besides, it has holes where other
regular orbits lie.

\begin{figure}
\centering
\includegraphics[width=7.6cm]{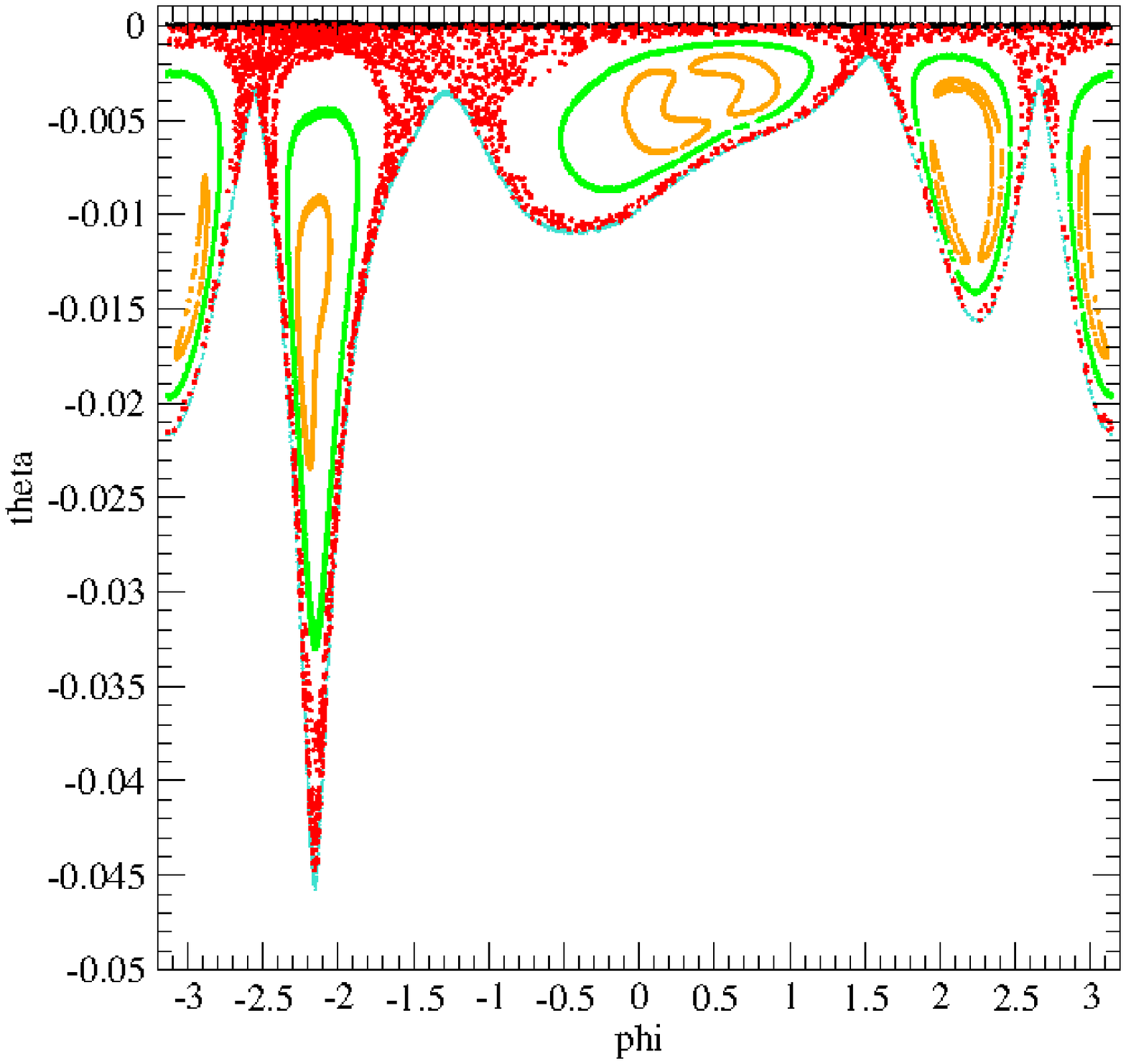}
\vskip 5.mm
\includegraphics[width=7.6cm]{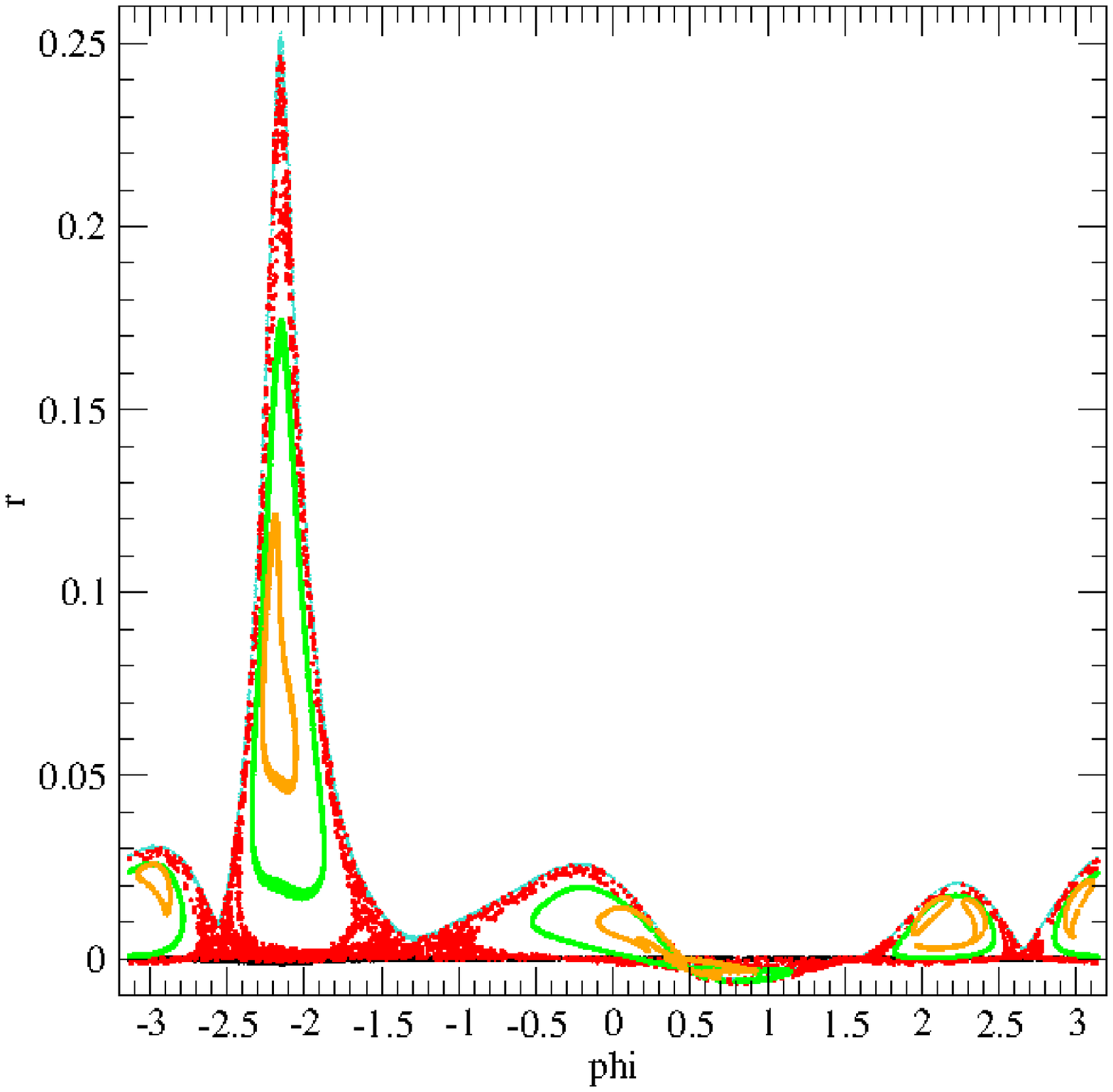}
    \caption{Same as Figure 14, but for the cuts $x = 0$ (u > 0) and
$\vert z \vert \leq 0.00005~(w>0)$.}
    \label{fig16}
\end{figure}
 
It is remarkable that, although we found the bounding regular orbits r1 and
r2 using only fits to very small sections of those orbits and of the partially
chaotic orbit pch, we find now a perfect match to the whole orbits, but there
is still more. Figure~\ref{fig15} is similar to the previous one, but it
corresponds to the left lobe, i.e., the one that had not been used at all to
find r1 and r2. Again, it is very similar to ordinary Poincar\'e maps, with
the partially chaotic orbit pch bounded by the regular orbits r1 and r2 and
with a hole that contains the regular orbits r3 and r4. Finally,
Figure~\ref{fig16} presents again results for the right lobe, but this time
they come from the cuts $x~(u > 0)$ and  $\vert z \vert \leq 0.00005~(w>0)$,
i.e., using $z$ instead of $y$ and, once again, different from the cuts used
to find r1 and r2. The conclusion that pch is a partially chaotic orbit
bounded by regular orbits r1 and r2 seems therefore unavoidable,
but it should be recalled that it is valid over time intervals of the order of
that covered by our numerical integrations. 
 
\section{Conclusions}

We have found a group of partially chaotic orbits inmersed in the mostly regular
domain around the double resonance (2,-1,0) and (0,1,-1) of the unperturbed
Hamiltonian~\ref{hamilt}. They are double orbits joined by a bifurcation which
is the likely cause of their chaotic nature. Those orbits, all with initial
conditions with $x = y = z = 0$, occupy a lane on the $(e_{1}, e_{2})$
plane, sections of which are shown on Figure~\ref{fig02}. That lane is,
therefore, made up of the 1-D curves that result from cutting 4-D partially
chaotic orbits with the planes $x = 0$, $y = 0$ and $z = 0$ and we could
identify one of those curves, corresponding to our orbit pch and shown on
Figure~\ref{fig08}. Moreover, a 4-D partially chaotic orbit can be bounded
by 3-D regular orbits, and we found two such bounding orbits, r1 and r2, in
Subsection~\ref{boundaries}. Finally, in Subsection~\ref{maps}, we presented
3-D Poincar\'e maps that: 1) Are very similar to the usual Poincar\'e maps
for autonomous Hamiltonians with two degrees of freedom; 2) Show that the
partially chaotic orbit pch is actually bounded by the regular orbits r1
and r2; 3) Show that, as in some 2-D cases, the partially chaotic orbit has
holes that contain additional regular orbits. In brief, we have made up a
strong case for the existence of partially chaotic orbits in cocoons well
isolated from the Arnold web by the regular orbits.
Our conclusions are supported by numerical integrations
and, therefore, are valid over time intervals of the order of those covered
by the integrations, but for elliptical galaxies those intervals are equivalent
to about one million Hubble times, i.e., much longer than the galactic lifetimes.

An important byproduct of our work is that, despite the difficulties to
extend to 3-D the use of the Poincar\'e maps, we have shown that it is
possible and, actually, 3-D Poincar\'e maps were a very important tool
for the present investigation.

We have already noted that other chains of probably partially chaotic orbits
can be seen in our Figure~\ref{fig01}, in the region of the double resonance
(2,-1,0) and (0,1,-1), so that the example shown here does not seem to be just
a rara avis. Moreover, we have found similar lanes of probably partially chaotic
orbits near other double resonances of the same Hamiltonian investigated here,
e.g., near $e_{1} = 0.35277369$ and $e_{2} = -0.34975917$ in the region of the
double resonance (2,-3,0) and (4,0,-3). Nevertheless, the structure of those
orbits is different from that of the simple croissants studied here and to
prove that they are actually partially chaotic will demand
a whole new investigation.

Of course, the big prize would be to show that partially chaotic orbits
bound fully chaotic ones, placing even more significant limits to chaotic
diffusion. But it is hopeless to try to do that by simply extending the
same techniques used here. In order to get a significant number of points,
our 3-D maps demand much longer integration times than the regular 2-D
ones and, to make one additional cut (say, $x = y = z = 0$), will require
prohibitely longer numerical integrations. Besides, there is another limitation:
we know from theory that regular orbits are tori and, therefore, that they
can bound partially chaotic orbits, but we do not know whether the latter
are closed surfaces or not and, therefore, we do not know whether they can
actually bound fully chaotic orbits. Nevertheless partially chaotic orbits,
having only one dimension less than fully chaotic ones, can at least pose
some barriers to the latter and, therefore, they might be a factor to be taken
into account in studies of chaotic diffusion. Therefore, we plan to continue
investigating this subject.

\section*{Acknowledgements}

We are very grateful to A. Jorba, M. Zou and D. Pfenniger for the use their codes,
and to D.D. Carpintero, R.E. Mart\'{\i}nez (sadly, recently deceased), M.E. Muzzio,
P. Santamar\'{\i}a, H.R. Viturro and F.C. Wachlin for their assistance. Special
thanks go to an anonymous reviewer whose suggestions were very useful to
improve the original version of the present paper. This work
was supported with grants from the Consejo Nacional de Investigaciones
Cient\'{\i}ficas y T\'ecnicas de la Rep\'ublica Argentina, the Agencia
Nacional de Promoci\'on Cient\'{\i}fica y Tecnol\'ogica and the Universidad
Nacional de La Plata.


\bsp	
\label{lastpage}
\end{document}